\documentclass[aip,jap,twocolumn,12pt,reprint]{revtex4-1}

\usepackage{amssymb}
\usepackage{amsmath}
\usepackage{graphicx}
\usepackage{dcolumn}
\usepackage{bm}

\begin{document}

\title{Magnetic detectability of a finite size paramagnet/superconductor cylindrical cloak}
\author{S.V.~Yampolskii}
\email[Electronic address: ]{yampolsk@mm.tu-darmstadt.de}
\author{Y.A.~Genenko}
\affiliation{Institut f\"{u}r Materialwissenschaft, Technische Universit\"{a}t Darmstadt,
Jovanka-Bontschits-Stra{\ss}e 2, D-64287 Darmstadt, Germany}
\date{\today }

\begin{abstract}
Cloaking of static magnetic fields by a finite thickness type-II superconductor tube surrounded by a 
coaxial paramagnet shell is studied. On the basis of exact solutions to the London and Maxwell equations, 
it is shown that perfect cloaking is realizable for arbitrary geometrical parameters including the thin 
film case for both constituents. In contrast to previous approximate studies assuming perfect diamagnetism 
of the superconductor constituent, it is proven that cloaking provides simultaneously full undetectability, 
that is the magnetic moment of the structure completely vanishes as well as all high-order multipole moments 
as soon as the uniform field outside remains unaffected. 
\end{abstract}

\maketitle

Electromagnetic metamaterials are currently of a great interest 
because they reveal many unusual, previously unrealizable properties~\cite{review2012}. One 
of the most intriguing effects is cloaking of electromagnetic waves as well as of static magnetic or electric 
fields~\cite{Science1,Science2,Science3,Pendry2007,NM2013,ElCloak,NM2008}.
A magnetic cloak is expected to produce a dual effect: it must not distort the external field 
outside the cloak, thus being "invisible" for external observation,  and, on the other hand, 
has to protect its inner area from the external field penetration. To realize these features different cloak 
designs have been proposed. Pendry et al. proposed~\cite{Science2} and Schurig et al.~\cite{Science3} 
experimentally realized a cloak for microwave electromagnetic fields 
using the (composite) material with anisotropic positionally dependent relative permeability and permittivity. 
Production of media and devices with spatially variable properties is thus possible but very 
complicated, that is why simpler hybrid systems consisting of ferromagnet and superconductor constituents 
were alternatively designed~\cite{NM2013,SuSTMawatari,APL102,SuST26,NJP13,Navau2009}. Recently, a magnetic 
cloak was experimentally realized in the forms of multilayered~\cite{AdvMater2012} or 
bilayered~\cite{Science2012,NJP15} magnet/superconductor hollow cylinder.

An essential component of the proposed hybrid cylindrical designs is the inner superconducting layer 
which was assumed to be an ideal diamagnetic medium with zero effective permeability in both 
analytical and finite-element considerations~\cite{NJP13,Science2012}. 
This assumption provides the cloaking effect ensuring that an external field does not penetrate
inside the cylinder and that the field of some magnetic source inside the cloak does not leak outside. 
However, it is intuitively clear that such idealization may have only a restricted validity.
Indeed, in reality magnetic field penetrates a superconductor to a finite depth even in the
Meissner state (the London penetration depth $\lambda$ in the case of bulk superconductors~\cite{deGennes}). 
In this respect following questions arise: Is perfect cloaking possible taking into account field
penetration in a superconductor? Can the cloak be made undetectable by magnetic measurements in this case?
Is it possible to completely protect the inner space of the cloak from the external field?

In the present Letter, we demonstrate the cloaking effect in a realistic cylindrical design of 
bilayer paramagnet/superconductor tube with finite thicknesses of both superconducting and magnetic 
constituents by exact solving the coupled London and Maxwell equations for respective media. 
We establish the values of constituent parameters necessary for perfect cloaking and prove the 
completely vanishing detectability of this object. This means that a perfect cloak of the considered 
paramagnet/superconductor cylindrical design inherently possesses properties of an "antimagnet" 
discussed in the literature~\cite{ElCloak,NJP13,SuST26,Mawa2012}.

\begin{figure}[!bp]
\includegraphics[width=8cm]{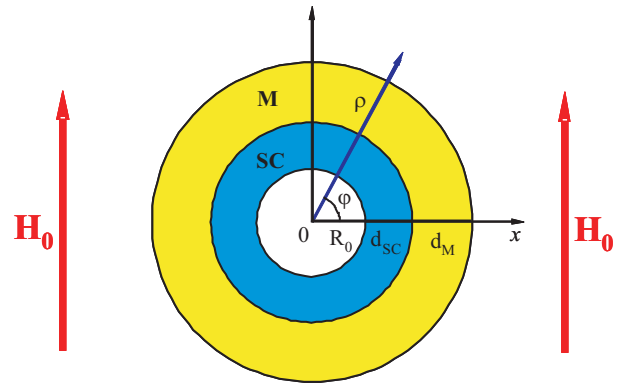}
\caption{(color online) Cross-sectional view of a hollow superconductor cylinder covered
by a coaxial cylindrical magnetic sheath.} \label{fig1}
\end{figure}

Let us consider an infinitely long hollow superconducting cylinder of thickness $d_S$ and radius of a 
coaxial hole $R_0$ enveloped in a coaxial cylindrical magnetic sheath of thickness $d_M$ with relative 
permeability $\mu $. This structure is exposed to an external constant magnetic field $\mathbf{H}_{0}$ 
perpendicular to the cylinder axis (Fig.~\ref{fig1}).

We assume that the superconductor layer is in the Meissner state so that 
the magnetic induction $\mathbf{B}_{S} $ in its region obeys the London 
equation~\cite{deGennes}%
\begin{equation}
\mathbf{B}_{S}+\lambda ^{2} \text{curl curl } \mathbf{B}_{S}=0.  
\label{1}
\end{equation}%
The magnetic field $\mathbf{H}_{in}$ inside the hole  
as well as the field outside the superconductor, denoted by $\mathbf{H}_{M}$ in a 
magnetic sheath and by $\mathbf{H}_{out}$ in a surrounding free space,
 are described by the Maxwell equations%
\begin{equation}
\text{curl } \mathbf{H}=0,\quad \text{div } \mathbf{B}=0,   \label{2}
\end{equation}%
the latter of which applies in the whole space. Implying an insulating,
nonmagnetic layer of thickness much less than $d_M$ and $d_{S}$ between the
superconductor and the magnet sheath, which is typical for hybrid magnet/superconductor 
structures (see, for example, Refs.~\cite{Lange2002,Dou2004}), 
the boundary conditions read%
\begin{subequations}
\label{BC}
\begin{align}
B_{S,n}& =\mu _{0} H_{in,n},\qquad  B_{S,t} =\mu _{0}H_{in,t}; 
\label{3a} \\
B_{S,n}& =\mu _{0}\mu H_{M,n},\quad B_{S,t} =\mu _{0}H_{M,t}; 
\label{3b} \\
\mu H_{M,n}& =H_{out,n},\quad \quad H_{M,t} =H_{out,t},   \label{3c}
\end{align}
\end{subequations}
for the normal ($n$) and tangential ($t$) components on the inner superconductor 
surface [Eq.~(\ref{3a})], on the superconductor/magnet interface [Eq.~(\ref{3b})] 
and on the outer magnet surface [Eq.~(\ref{3c})], respectively (cf. Refs.~\cite{APL2004,PRB2005}),
with $\mu_0$ permeability of vacuum.
In addition, the field $\mathbf{H}_{out}$ has to approach asymptotically the 
external field $\mathbf{H}_{0}$.

In cylindrical coordinates ($\rho ,\varphi ,z$) coaxial with the tube
the solution of Eqs.~(\ref{1})-(\ref{2}) is
\begin{eqnarray}
H_{in,\rho} &=&H_{0}A_{in}\sin \varphi ,  
\label{Hhole} \\
H_{in,\varphi} &=&H_{0}A_{in} \cos \varphi   \notag
\end{eqnarray}
in the hole ($ \rho \le R_0$),
\begin{eqnarray}
B_{S,\rho} &=&\mu _{0} H_{0} \left\{ A_{S1} \left[ I_{0}\left( \rho /\lambda \right)
-I_{2}\left( \rho /\lambda \right) \right] \right.  \label{HSC}\\  
&+& \left. A_{S2}\left[ K_{0}\left( \rho /\lambda \right)
-K_{2}\left( \rho /\lambda \right) \right] \right\} \sin \varphi ,  \nonumber
 \\
B_{S,\varphi} &=&\mu _{0}H_{0}\left\{ A_{S1}\left[ I_{0}\left( \rho /\lambda
\right) +I_{2}\left( \rho /\lambda \right) \right] \right. \notag \\  &+& \left. A_{S2}\left[ K_{0}\left( \rho /\lambda
\right) +K_{2}\left( \rho /\lambda \right) \right] \right\} \cos \varphi   \notag
\end{eqnarray}
in the superconductor ($ R_0 \le \rho \le R_1= R_0 + d_S$),
\begin{eqnarray}
H_{M,\rho} &=&H_{0}\left( A_{M1}-A_{M2}R^{2}_1/\rho ^{2}\right) \sin \varphi ,
  \label{HML} \\
H_{M,\varphi} &=&H_{0}\left( A_{M1}+A_{M2}R^{2}_1/\rho ^{2}\right) \cos
\varphi   \notag
\end{eqnarray}%
in the magnet sheath ($ R_1 \le \rho \le R_2= R_1 + d_M$), and
\begin{eqnarray}
H_{out,\rho} &=&H_{0}\left( 1+A_{out} R_2^2/\rho ^{2}\right)
\sin \varphi ,   \label{Hspace} \\
H_{out,\varphi} &=&H_{0}\left( 1-A_{out} R_2^2/\rho ^{2}%
\right) \cos \varphi   \notag
\end{eqnarray}%
in the space around the tube ($ \rho \ge R_2$). The coefficients $A_{in}$, $A_{S1}$, $A_{S2}$, $A_{M1}$, $A_{M2}$ and 
$A_{out}$ determined from the boundary conditions~(\ref{BC}) are quite cumbersome and, therefore, 
given in Appendix A.

In order to leave the magnetic field outside the cloak undisturbed, a condition $A_{out} = 0$ has to be fulfilled.
This results in the following equation:
\begin{widetext}
\begin{equation}
\frac{\left( \mu + 1 \right)^2 - \left( \mu -1 \right)^2 R_2^2 / R_1^2}
{\left( \mu^2 -1 \right) \left( R_2^2 /R_1^2 -1 \right)} 
=  \frac{I_0 \left( R_1/\lambda \right) K_2 \left( R_0/\lambda \right)- 
K_0 \left( R_1/\lambda \right) I_2 \left( R_0/\lambda \right)}
{I_2 \left( R_1/\lambda \right) K_2 \left( R_0/\lambda \right)- 
K_2 \left( R_1/\lambda \right) I_2 \left( R_0/\lambda \right)},  \label{CloakExt}
\end{equation}
\end{widetext}
\noindent from which a magnitude of the relative permeability $\mu \left( d_M \right) $ 
can be found which provides the cloaking effect.

\begin{figure}[!tbp]
\includegraphics[width=8cm]{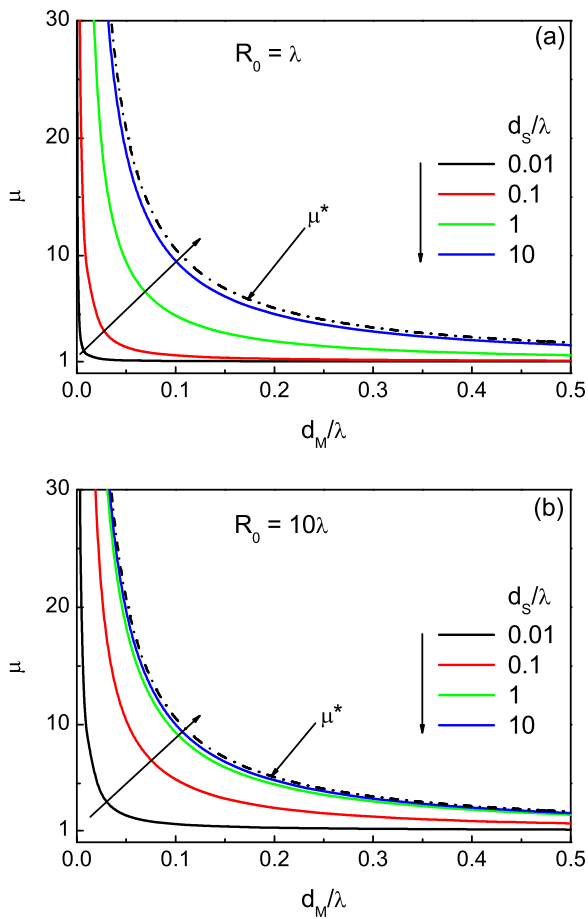}
\caption{(color online) Solutions of Eq.~(\ref{CloakExt}) for different values of the superconductor layer thickness $d_S$ 
and of the cylinder hole radius $R_0$: a) $R_0 = \lambda$; b) $R_0 = 10 \lambda$. 
In both plots the dash-dotted curve depicts the idealized dependence $\mu^* \left( d_M \right)$ 
of Eq.~(\ref{mu_ideal}).}
\label{fig2}
\end{figure}

In Fig.~\ref{fig2} the dependences of $\mu$ on the thickness $d_M$ of the magnet sheath 
are shown for different thicknesses $d_S$ of the superconductor layer 
in the cases of small ($R_0 = \lambda$) and large ($R_0 = 10 \lambda$) inner hole of the cloak. 
One can see that in both cases the cloaking effect 
exists for wide range of parameters of the magnet layer which, in general, are different from the values 
\begin{equation}
\mu^*  = \left( R_2^2 + R_1^2 \right) /\left( R_2^2 - R_1^2 \right) \label{mu_ideal}
\end{equation}
obtained earlier for the cloak with idealized superconductor layer (viz., considered as a layer with 
zero magnetic permeability)~\cite{Science2012}. 

It is possible to derive some approximate solutions of Eq.~(\ref{CloakExt}). 
In the case of a macroscopic cloak, i.e., if $R_0 \gg \lambda$, its right-hand side 
is simplified to the form  $1+\left(2 \lambda / R_1 \right) \coth \left( d_S / \lambda \right)$
and in the case of thick ($d_S \gg \lambda$) 
superconductor layer we then obtain
\begin{equation}
\mu = \frac{R_2^2 + R_1^2}{R_2^2- R_1^2} \times \left[ 1- \frac{4 \lambda}{R_1} 
\frac{R_2^2 R_1^2}{\left( R_1^2 + R_2^2 \right)^2} \right].  \label{mu_thick}
\end{equation}
\noindent If the superconductor layer is thin ($d_S \ll \lambda$) but $R_1 \gg \Lambda = \lambda^2 /d_S$ 
($\Lambda$ is Pearl's effective penetration depth~\cite{deGennes,Pearl}), the permeability $\mu$ is also described by 
Eq.~(\ref{mu_thick}) with the only change $\lambda \to \Lambda$. 
And finally, expression (\ref{mu_thick}) applies also in the case of a thick superconductor 
with $d_S \gg \lambda$ for the inner hole size $R_0 \ll \lambda $. 
Notice that the main term of expression (\ref{mu_thick}) coincides with $\mu^*$ and, therefore, 
the applicability of the idealized dependence~(\ref{mu_ideal}) in the macroscopic case is ensured. 
Additionally, for both thin and thick superconductors the cloaking effect can be provided already with rather 
thin magnet sheath with $d_M\ll R_1$ if the relative permeability satisfies the condition
$\mu d_M/R_1\simeq 1$. This is in agreement with the known effectiveness of magnetic shielding as soon as 
the strength of an effective magnetic dipole layer $\mu d_M/R_1$ is notable~\cite{GenenkoJAP2002,GenenkoPhysC2004-1}.

As for the possible screening of the inner space of the cloak from the external magnetic field, 
by using of Eq.~(\ref{CloakExt}) it follows that 
\begin{equation}
A_{in} = \frac{\left( \mu^2 -1 \right) \left( R_2^2/R_1^2 -1 \right)}
{ 2 \mu \left[ I_2 \left( R_1/\lambda \right) K_2 \left( R_0/\lambda \right)- 
K_2 \left( R_1/\lambda \right) I_2 \left( R_0/\lambda \right) \right]}, \label{AinCloak}
\end{equation}
\noindent provided that $A_{out} = 0$. Therefore, a non-zero homogeneous magnetic field $H_{y,in} = H_0 A_{in}$ 
prevails inside the cloak. In the case of a macroscopic cloak with $R_0 \gg \lambda, \Lambda$ this field, 
in the main approximation, reads
\begin{equation}
H_{y,in} = H_0 \frac{\left( R_0 R_1 \right)^{1/2}}{\lambda} \frac{\left( \mu^2 -1 \right)}{2 \mu} 
\frac{\left( R_2^2/R_1^2 -1 \right)}{\sinh \left( d_S / \lambda \right)}, \label{Hy_macro}
\end{equation}
decreasing exponentially only in the case of $d_S \gg \lambda$ and tending to a more pronounced dependence
on the magnet layer parameters in the opposite case of thin superconductors.

An important question concerning the "invisibility" of the coaxial bilayer structure is whether or not 
it can be detected~\cite{NJP13}, for example, by measuring a magnetic moment of this structure. 
Due to the geometry of the system, this moment per unit length of the structure has only a $y$ component and consists of two parts:
the magnetic moment of the superconductor layer defined by means of the Meissner current density 
$\mathbf j$ as~\cite{APL2004}
\begin{equation}
M_S = \int_{V_S} dV \left[ \bm{\rho} \times \mathbf{j} \right]_y,
\end{equation}
and the moment of magnetic sheath defined as~\cite{APL2004}
\begin{equation}
M_M =  \left( \mu -1 \right) \int_{V_M} dV H_{M,y}.
\end{equation}
By calculating these contributions (for details, see Appendix B), 
one finds that the total magnetic moment per unit length of the structure $M=M_S+M_M$ is proportional to the coefficient $A_{out}$,
\begin{equation}
M = 2 \pi R_2^2 H_0 A_{out}.
\end{equation}
This means that in the cloaking case ($A_{out}=0$) the magnetic moment of the system simultaneously vanishes 
ensuring that the object cannot be detected by magnetic measurements. Indeed, according to the form of 
the solution~(\ref{Hhole})-(\ref{Hspace}) the structure under consideration does not possess other
multipole moments but the dipole one. If the latter equals zero the object cannot be observed by any magnetic
measurement, at least as long as an external field uniform at the scale of the object transverse size 
$\sim R_2$ is involved.

The above considered flux-free state of the superconductor is restricted to relatively small magnetic fields. 
Within the London theory this state is protected by the Bean-Livingston barrier enhanced by the magnetic 
sheath~\cite{GenenkoJPCM2005,PRB2005} which prevents the first magnetic vortex penetration below a 
characterictic field
\begin{equation}
\label{ctritH}
H_p = \left( \Phi_0 / 16\pi\mu_0\lambda \sigma \right) \ln{(e\sigma/\xi)} 
\left[ \mu +1 -(\mu -1)R_1^2/R_2^2  \right]
\end{equation}
\noindent where $\Phi_0$ is the magnetic flux quantum, $\xi$ the superconductor coherence length and 
$\sigma\in (\xi,\lambda)$ the typical defect size at the superconductor surface. 
On the other hand, for the case of multiple vortex penetration, the limiting field can be estimated 
within the critical state model~\cite{Mawa&Clem2006} to be of the 
order of $j_c d_S$ with $j_c$ the critical current density in the superconductor. The above consideration 
of cloaking is valid until the highest of these two fields is exceeded. This does not preclude, however, 
the possibility of cloaking when a superconductor is in the critical state, but this case needs a more
elaborated treatment. 

An interesting intermediate case is that of an external magnetic field slightly exceeding the penetration field 
(\ref{ctritH}), $ H_0 \ge H_p$, when magnetic flux may penetrate the superconductor layer in the form of a single
vortex loop~\cite{PRB2005} and possibly be trapped there. Such a vortex will add a paramagnetic moment $M_V$ which, 
in general, will break the perfect cloaking. The approximate maximum value of the moment $M_V$ for a single vortex 
may be estimated in the limit of $R_1 \gg \lambda$ under cloaking conditions (for details, see Appendix C). 
Comparing this value with the absolute value of the magnetic moment per unit length of 
the superconductor constituent, one finds that
\begin{equation}
\frac{M_V}{| M_S | L} \simeq \frac{H_p^0}{8 H_0} \frac{d_S^2}{R_1^2}\frac{\sigma}{L \ln (e \sigma / \xi)} 
\frac{R^4_2 + R_1^2 R_2^2 + 2 R_1^4}{R_2^2 \left( R_2^2 + R_1^2 \right)},
\end{equation}
\noindent where $L$ is the length of the sample. Such an estimate shows that, though the total magnetic moment of the 
system does not vanish, it remains rather small due to the factor $\sigma / L $ for a thick superconductor layer
and due to the additional small factor $( d_S / R_1 )^2$ for the case of a thin superconductor layer.

In conclusion, we have studied theoretically static magnetic cloaking properties of a realistic bilayer 
paramagnet/superconductor cylindrical tube with respect to the influence of both geometrical and 
material parameters of the system constituents. We have found that a non-distorted uniform magnetic field 
outside the cloak can exist in a wide range of relative permeability and thickness values of the magnet 
sheath for both cases of thick and thin superconductor layers. Under the above cloaking conditions the
magnetic moment of the bilayer structure vanishes (as well as all higher multipole moments) making this
object magnetically undetectable. Thus, as soon as a uniform external magnetic field is concerned there is 
no difference between a perfect cloak and an "antimagnet" introduced in Ref.~\cite{NJP13}. In the case of 
an arbitrary nonuniform external field this subtle difference might exist so that this question needs more
sophisticated treatment.
Another restriction on cloaking results from the finite thickness of 
superconducting constituent. Due to this property such a system never completely protects the inner region 
(a central hole) from the penetration of the external field. Therefore, we suppose that using such a cloak 
design for simultaneous protection of sensitive equipment, proposed in the 
literature~\cite{NJP13,NJP15,Science2012}, requires further consideration. 

\appendix
\section{Calculation of the magnetic field distribution}

In cylindrical coordinates ($\rho ,\varphi ,z$) coaxial with the tube
the distribution of magnetic field in the superconductor layer is found by direct solving of the London equation~(\ref{1})
together with the condition $\text{div } \mathbf{B}_S = 0$. In the other regions of the system it is convenient
to represent the magnetic induction in the terms of a vector potential, $\mathbf{B} = \text{curl } \mathbf{A}$.
In the considered geometry, accounting for the gauge invariance of the vector potential 
$\text{div } \mathbf{A} = 0 $, it suffices to consider only the component $A_z \left( \rho, \varphi \right) $  
which should satisfy the equation
\begin{equation}
\bm{\nabla}^2 A_z \left( \rho, \varphi \right) = 0. \label{Az} 
\end{equation}
The general solution of Eq.~(\ref{Az}) is
\begin{equation}
A_z \left( \rho, \varphi \right) = \sum_m \exp \left( i m \varphi \right) \left( C_{1m} \rho^m +
C_{2m} \rho^{-m} \right).  \label{AzSol1} 
\end{equation}
The magnetic field outside the system, $\mathbf{H}_{out}$, at large distances $\rho \to \infty$ 
should approach the homogeneous external field having in cylindrical coordinates the components 
$H_{0,\rho} = H_0 \sin \varphi$ and $H_{0,\varphi} = H_0 \cos \varphi$. Because of this, 
only the terms with $m = \pm 1$ in the expansion~(\ref{AzSol1}) provide a non-trivial solution of 
the system of equations following from the boundary conditions~(3).

Finally, the distribution of magnetic field is described by the formulas~(\ref{Hhole})-(\ref{Hspace}), where
the coefficients $A_{in}$, $A_{S1}$, $A_{S2}$, $A_{M1}$, $A_{M2}$ and 
$A_{out}$ read
\begin{eqnarray}
A_{in} & = & \left( 8\mu /\Delta \right) \left(\lambda /R_0 \right)^2, \\   \label{Ain} 
A_{S1} & = & 4 \mu K_2 \left( R_0 / \lambda \right) / \Delta, \label{AS1}  \\     
A_{S2} & = & - 4 \mu I_2 \left( R_0 / \lambda \right) / \Delta,  \label{AS2}   
\end{eqnarray}
\begin{widetext}
\begin{eqnarray}
A_{M1} & = & 2 \left\{ \left[I_0 \left( R_1/\lambda \right) K_2 \left( R_0/\lambda \right)- 
K_0 \left( R_1/\lambda \right) I_2 \left( R_0/\lambda \right) \right] \left( \mu +1 \right) \right. \label{AM1} \\
& + & \left. \left[I_2 \left( R_1/\lambda \right) K_2 \left( R_0/\lambda \right)- 
K_2 \left( R_1/\lambda \right) I_2 \left( R_0/\lambda \right) \right] \left( \mu -1 \right)\right\} /
 \Delta,  \nonumber \\
A_{M2} & = & 2 \left\{ \left[I_0 \left( R_1/\lambda \right) K_2 \left( R_0/\lambda \right)- 
K_0 \left( R_1/\lambda \right) I_2 \left( R_0/\lambda \right) \right] \left( \mu -1 \right) \right.  \label{AM2} \\
& + & \left. \left[I_2 \left( R_1/\lambda \right) K_2 \left( R_0/\lambda \right)- 
K_2 \left( R_1/\lambda \right) I_2 \left( R_0/\lambda \right) \right] \left( \mu +1 \right)\right\} /
 \Delta, \nonumber \\
A_{out} &=&\left\{ \left[I_0 \left( R_1/\lambda \right) K_2 \left( R_0/\lambda \right)- 
K_0 \left( R_1/\lambda \right) I_2 \left( R_0/\lambda \right) \right] 
\left( \mu ^{2}-1\right) \left( 1-R^{2}_1/ R^{2}_2\right)  \right. \label{Aout} \\
&+& \left. \left[I_2 \left( R_1/\lambda \right) K_2 \left( R_0/\lambda \right)- 
K_2 \left( R_1/\lambda \right) I_2 \left( R_0/\lambda \right) \right] 
\left[ \left( \mu -1\right) ^{2}-\left( \mu +1\right)
^{2}R_1^{2}/R_2^{2}\right] \right\} / \Delta,  \nonumber 
\end{eqnarray}
with
\begin{eqnarray}
\Delta &=&\left[I_0 \left( R_1/\lambda \right) K_2 \left( R_0/\lambda \right)- 
K_0 \left( R_1/\lambda \right) I_2 \left( R_0/\lambda \right) \right] 
\left[ \left( \mu +1\right) ^{2}-\left( \mu -1\right)
^{2}R_1^{2}/R_2^{2}\right] \label{det} \\
&+& \left[I_2 \left( R_1/\lambda \right) K_2 \left( R_0/\lambda \right)- 
K_2 \left( R_1/\lambda \right) I_2 \left( R_0/\lambda \right) \right] 
\left( \mu ^{2}-1\right) \left( 1-R^{2}_1/ R^{2}_2\right).  \nonumber
\end{eqnarray}

\section{Magnetic moment of the structure in the Meissner state}

The magnetic moment of the superconductor per unit length along the cylinder axis is defined as~\cite{APL2004}
\begin{equation}
M_S = \int_{V_S} dV \left[ \bm{\rho} \times \mathbf{j} \right]_y, \label{MS}
\end{equation}
where the Meissner current density $\mathbf j$ has only a $z$ component which equals
\begin{equation}
j_z \left( \rho, \varphi \right) = \frac{1}{\mu_0 \rho} \left[ \frac{\partial}{\partial \rho} 
\left( \rho B_{S,\varphi} \right)- \frac{\partial B_{S,\rho}}{\partial \varphi} \right] = \left( 2 H_0 / \lambda \right) \cos \varphi \left[ 
A_{S1} I_1 \left( \rho / \lambda \right) - A_{S2} K_1 \left( \rho / \lambda \right)\right]. 
\end{equation}
\noindent After integration in Eq.~(\ref{MS}) one obtains
\begin{eqnarray}
M_S &=& -2 \pi H_0 \left. \left\{ \rho^2 \left[ A_{S1} I_2 \left( \rho / \lambda \right) +
A_{S2} K_2 \left( \rho / \lambda \right) \right] \right\} \right|_{R_0}^{R_1} \nonumber \\
 &=& -8 \pi H_0 R_1^2 \mu \left[ I_2 \left(R_1/\lambda \right) K_2 \left(R_0/\lambda \right) - 
K_2 \left(R_1/\lambda \right) I_2 \left(R_0/\lambda \right) \right] / \Delta.
\end{eqnarray}

The moment of magnetic sheath per unit length along the cylinder axis is defined as~\cite{APL2004} 
\begin{equation}
M_M =  \left( \mu -1 \right) \int_{V_M} dV H_{M,y} = \left( \mu -1 \right) 
\int_{V_M} dV \left(H_{M,\rho} \sin \varphi + H_{M,\varphi} \cos \varphi \right) \label{MM}
\end{equation}
and reads
\begin{eqnarray}
M_M &=& \pi \left( R_2^2 - R_1^2 \right) \left( \mu -1 \right) H_0 A_{M1} \nonumber\\
&=& 2 \pi \left( R_2^2 - R_1^2 \right) H_0 \left\{ \left[I_0 \left( R_1/\lambda \right) K_2 \left( R_0/\lambda \right)- 
K_0 \left( R_1/\lambda \right) I_2 \left( R_0/\lambda \right) \right] \left( \mu^2 -1 \right) \right. \nonumber\\
& + & \left. \left[I_2 \left( R_1/\lambda \right) K_2 \left( R_0/\lambda \right)- 
K_2 \left( R_1/\lambda \right) I_2 \left( R_0/\lambda \right) \right] \left( \mu -1 \right)^2 \right\} /
 \Delta. 
\end{eqnarray}
\end{widetext}
It is easy to obtain that the total magnetic moment $M=M_S+M_M$ of the paramagnet/superconductor tube per unit length is 
\begin{equation}
M = 2 \pi R_2^2 H_0 A_{out},
\end{equation}
\noindent identifying simultaneously perfect cloaking and magnetic undetectability of the system.
In the cloaking case, by using the condition $A_{out}=0$ (i.e., Eq.~(\ref{CloakExt})) the absolute value 
of the superconductor moment $M_S$ (and the same for the magnet shield moment $M_M$) can be reduced to the form 
\begin{equation}
|M_S| = \frac{\pi H_0 R_1^2}{2 \mu} \left( \mu^2 -1 \right)\left( \frac{R_2^2}{R_1^2}-1\right),  \label{MSCloak}
\end{equation}
with permeability $\mu$ satisfying Eq.~(\ref{CloakExt}).

\section{Magnetic moment of a single vortex loop at cloaking conditions}

It is known that, at a flat paramagnet/superconductor boundary, magnetic flux penetrates the superconductor
in the form of a small vortex loop~\cite{GenenkoJPCM2005} when the transverse external field exceeds some 
characteristic field $H_p^0 =  \left( \Phi_0 / 4\pi\mu_0\lambda \sigma \right) \ln{(e\sigma/\xi)}$, here 
$\Phi_0$ is the magnetic flux quantum, $\xi$ the superconductor coherence length and 
$\sigma\in (\xi,\lambda)$ the typical defect size at the superconductor surface. 
In the cylindrical structure under consideration the first vortex loop of radius $r$ penetrates the superconductor
in the field $H_p = H_p^0 \left[ \mu +1 -(\mu -1)R_1^2/R_2^2  \right]/4$ (Ref.~\cite{PRB2005}) and, neglecting the 
effect of the inner hole when $r\ll d_S$, its magnetic moment may be described by the expression~\cite{PRB2005} 
\begin{widetext}
\begin{equation}
M_V = M_V^0 \frac{2 \mu + \left( \mu^2 + 1\right)\left(R_2^2/R_1^2 -1 \right)}
{2 \mu + \left( \mu + 1\right)\left(R_2^2/R_1^2 -1 \right) \left[1 + \left( \mu - 1 \right) 
I_1^\prime \left(R_1/ \lambda \right) / I_0 \left(R_1/ \lambda \right)\right]},
\end{equation}
\end{widetext}
\noindent where the magnetic moment of the loop in a magnetically unshielded superconductor is~\cite{PRB57}
\begin{equation}
M_V^0 \simeq \frac{\Phi_0 r^2}{4 \mu_0\lambda  } \frac{I_1 \left(R_1/ \lambda \right) }{I_0 \left(R_1/ \lambda \right)}.
\end{equation}
In the limit of $R_1 \gg \lambda$, the moment $M_V$ can be estimated as follows
\begin{equation}
M_V = M_V^0 \frac{2 \mu + \left( \mu^2 + 1\right)\left(R_2^2/R_1^2 -1 \right)}
{\mu \left[ 2 + \left( \mu + 1\right)\left(R_2^2/R_1^2 -1 \right) \right]}
\end{equation}
with $M_V^0 = \Phi_0 r^2/4 \mu_0\lambda$. At cloaking conditions, substituting 
$\mu = \mu^*$ from Eq.~(\ref{mu_ideal}) and
assuming characteristic radius of the vortex loop $r \lesssim d_S$, in the main approximation one obtains
\begin{equation}
M_V = \frac{\Phi_0 d_S^2}{4 \mu_0\lambda}  \frac{R_2^4 + R_1^2 R_2^2 +2 R_1^4}{\left(R_2^2 + R_1^2 \right)^2}.
\end{equation}
Using expression~(\ref{MSCloak}) with $\mu = \mu^*$, one can easily obtain that at the cloaking condition
\begin{equation}
\frac{M_V}{| M_S | L} \simeq \frac{H_p^0}{8 H_0} \frac{d_S^2}{R_1^2}\frac{\sigma}{L \ln{(e \sigma / \xi)}} 
\frac{R^4_2 + R_1^2 R_2^2 + 2 R_1^4}{R_2^2 \left( R_2^2 + R_1^2 \right)},
\end{equation}
where $L$ is the length of the superconductor constituent.

\vspace{5mm}
\bibliographystyle{plain}
\bibliography{apssamp}

\end{document}